\documentclass[conference]{IEEEtran}
\IEEEoverridecommandlockouts

\usepackage{cite}
\usepackage{amsmath,amssymb,amsfonts}
\usepackage{algorithmic}
\usepackage{graphicx}
\usepackage{url}
\usepackage{textcomp}
\usepackage{xcolor}
\usepackage{colortbl}
\usepackage{longtable}
\usepackage{xltabular}
\usepackage{threeparttable}
\usepackage[most]{tcolorbox}
\def\BibTeX{{\rm B\kern-.05em{\sc i\kern-.025em b}\kern-.08em
    T\kern-.1667em\lower.7ex\hbox{E}\kern-.125emX}}

\begin{document}

\title{Differences between Neurodivergent and Neurotypical Software Engineers: Analyzing the 2022 Stack Overflow Survey}

\author{\IEEEauthorblockN{Pragya Verma}
\IEEEauthorblockA{\textit{Reykjavik University} \\
Reykjavik, Iceland \\
pragyav@ru.is}
\and
\IEEEauthorblockN{Marcos Vinicius Cruz}
\IEEEauthorblockA{\textit{Reykjavik University} \\
Reykjavik, Iceland \\
marcosc@ru.is}
\and
\IEEEauthorblockN{Grischa Liebel}
\IEEEauthorblockA{\textit{Reykjavik University} \\
Reykjavik, Iceland \\
grischal@ru.is}
}

\maketitle

\begin{abstract}
Neurodiversity describes variation in brain function among people, including common conditions such as Autism spectrum disorder (ASD), Attention deficit hyperactivity disorder (ADHD), and dyslexia.
While Software Engineering (SE) literature has started to explore the experiences of neurodivergent software engineers, there is a lack of research that compares their challenges to those of neurotypical software engineers.
To address this gap, we analyze existing data from the 2022 Stack Overflow Developer survey that collected data on neurodiversity.
We quantitatively compare the answers of professional engineers with ASD (n=374), ADHD (n=1305), and dyslexia (n=363) with neurotypical engineers.
%
%
Our findings indicate that neurodivergent engineers face more difficulties than neurotypical engineers.
Specifically, engineers with ADHD report that they face more interruptions caused by waiting for answers, and that they less frequently interact with individuals outside their team.
This study provides a baseline for future research comparing neurodivergent engineers with neurotypical ones.
Several factors in the Stack Overflow survey and in our analysis are likely to lead to conservative estimates of the actual effects between neurodivergent and neurotypical engineers, e.g., the effects of the COVID-19 pandemic and our focus on employed professionals.
\end{abstract}




\section{Introduction}
Neurodiversity describes variation in brain function among people \cite{doyle2021diamond}, including conditions such as Autism spectrum disorder (ASD), Attention deficit hyperactivity disorder (ADHD), and dyslexia.
The neurodiversity movement argues that this variation should be seen as natural diversity among individuals and that society should accommodate this diversity, instead of focusing on finding a ``cure'' for people with disorders in the medical sense \cite{singer1998odd}.
Including approximately 15\% to 20\% of the world population~\cite{doyle2020neurodiversity}, neurodivergent (ND) individuals represent a major part of the actual and potential work force.

Software Engineering (SE) literature has started to explore the experiences of ND engineers for various conditions, e.g., for ADHD~\cite{liebel24_adhd}, ASD~\cite{marquez2024inclusion},  dyslexia~\cite{mcchesney18}, and combinations of these conditions \cite{gama25_gt_asd_adhd,morris2015understanding}.
However, there are only few studies in SE providing comparisons to neurotypical (NT) engineers, i.e., those that have what is considered 'normal' cognitive function.
As challenges and strengths of ND engineers overlap with those of NT engineers, and as most conditions included in ND exhibit large variation in symptoms, it is therefore unclear if and to what extent ND engineers differ from the general SE population.

To address this gap, we analyze existing data from the 2022 Stack Overflow Developer survey\footnote{\url{https://survey.stackoverflow.co/2022}}.
This survey has been conducted every year since 2011 by Stack Overflow among thousands of software developers, and collected data on neurodiversity for the first and last time in 2022.
We quantitatively compare the answers of professional developers with ASD (n=374), ADHD (n=1305), and dyslexia (n=363) with NT professionals.
We aim to answer the following research question (RQ):
\begin{description}
\item[\textbf{RQ:}] What differences exist between professional developers with dyslexia, ADHD, and ASD and their neurotypical peers?
\end{description}
Specifically, we compare the answers with respect to knowledge sharing, interaction, and finding information of each of these three groups with the same amount of NT developers, sampled randomly as well as sampled based on the work mode and the work experience distributions.

We find that ND developers, especially those with ADHD working in a hybrid work mode, face more difficulties in obtaining and sharing knowledge within their organization.
However, we find only few significant differences, and with small effect sizes.
Our findings indicate that difficulties with respect to the analyzed factors exist among ND professionals, but are not substantially different to NT professionals.
Various aspects are not considered in the Stack Overflow survey and might confound the results, e.g., the impact of accommodations at the workplace and tailoring of remote or hybrid work environments.
As such, our study likely provides a conservative baseline that can be used in future studies to further explore the challenges of ND professionals in SE, as well as how to address them.

The filtered and preprocessed data, as well as the corresponding scripts are available on Zenodo~\cite{dataset_SO}.

\section{Related Work}
Neurodiversity has emerged as a studied topic in SE literature in the last years, e.g., \cite{morris2015understanding,marquez2024inclusion,liebel24_adhd,mcchesney18,mcchesney20}.

To our knowledge, the first study on this topic was conducted by Morris et al.~\cite{morris2015understanding}, who investigate challenges faced by ND engineers in SE by conducting interviews with 10 ND engineers, followed by a survey among 846 engineers.
The authors find that ND engineers face various challenges related to their work and to interpersonal aspects, and that they fear stigmatization.
Several aspects relate specifically to SE, e.g., a tendency to get bored with mundane tasks, or expressing inappropriate emotions, e.g., when criticized during code reviews.
Their survey reveals some differences of ND engineers compared to their NT colleagues, namely that they find it comparably more challenging to work in shared offices, or deciding when to seek help for tasks.

In a series of qualitative studies based on interviews, Gama et al.~\cite{liebel24_adhd,gama25_gt_asd_adhd} investigate the experiences of engineers with ASD and ADHD in SE.
These studies find that the interviewed individuals face various challenges related to cognitive and emotional dysfunctions and to social interaction and communication.
In turn, these challenges lead to stress that affects their work performance. They suggest various accommodations that can help addressing the challenges and the resulting stress.
However, these three studies provide little insights in how the reported challenges, stressors, and accommodations compare to experiences by NT colleagues.

Focusing on dyslexia, a series of experiments by McChesney and Bond~\cite{mcchesney18,mcchesney20} use eye tracking to study how developers with dyslexia read software code.
They find that the studied individuals, in reading code, are not affected in the same way as when reading text.
Potential explanations are that reading code is significantly different from regular text, e.g., due to indentation or spacing.

Marquez et al.~\cite{marquez2024inclusion} conduct a systematic literature review on the inclusion of individuals with ASD in SE. They find that barriers are commonly report, but potential interventions are not studied empirically. The articles included in the review are primarily studies focused on education or secondary studies.  

Beyond SE, there are various studies on the practices and performance of ND individuals at the workplace, e.g., \cite{das2021towards,zolyomi2019managing,alqahtani2019understanding}
Das et al.~\cite{das2021towards} investigate the remote work practices of ND individuals during COVID-19, including also software developers.
The authors report that ND individuals tailor their home environments, successfully negotiate their communication practices with team members, and balance various tensions related to productivity and fatigue.

Zolyomi et al.~\cite{zolyomi2019managing} investigate the needs of individuals with ASD related to video calls.
The authors find their interviewees develop various coping mechanisms, such as adopting NT behavior.

Alqahtani et al.~\cite{alqahtani2019understanding} conduct an eye-tracking experiment on information processing performance with 24 participants, 12 with ADHD and 12 without.
The results show that participants exhibit similar performance when processing information in textual, tabular, or graphical form, but that preferences differ between the groups.
That is, participants with ADHD tend to rate the graphical representation lowest.

Based on a community and a clinical sample, Fuermaier et al.~\cite{fuermaier2021adhd} analyze the work performance of individuals with ADHD.
They find that individuals with ADHD especially struggle with not meeting their own expectations and potential, but that this does not commonly result in low evaluations.
Similarly, symptoms of inattention are strongly associated with work performance.

Rogers et al.~\cite{rogers2017fatigue} conduct a cross-sectional between-subjects study of individuals with and without ADHD.
Based on self-reported measures, the study finds that individuals with ADHD are significantly more fatigued than the control.
In a similar direction, based on perceived stress data collected from 983 individuals, Combs et al.~\cite{combs15} find that ADHD symptoms are positively associated with stress.
Finally, a cross-sectional study based on over 15.000 participants finds that ADHD symptoms are positively associated with workaholism \cite{andreassen2016relationships}.

Summarizing work on employment and employment-related challenges of individuals with ASD, Hendricks~\cite{hendricks2010employment} reports that interaction-related challenges are most common among this group.
Similarly, challenges related to executive function, e.g., difficulties in attention or working memory, hinder individuals with ASD.
Finally, similarly to individuals with ADHD, high levels of stress and anxiety are reported.

Finally, in a literature review on employment of individuals with Dyslexia, De Beer et al.~\cite{de2014factors} summarize 33 qualitative and quantitative studies according to the two-level classification of the International Classification of Functioning, Disability and Health\footnote{\url{https://www.who.int/standards/classifications/international-classification-of-functioning-disability-and-health}} (ICF).
The study finds that 318 of the ICF factors are covered in existing studies, with the most common ones related to mental functions (e.g., anxiety or frustration), activities (e.g., difficulties in reading and writing, but also strengths in problem solving and speaking), working conditions (e.g., positive and negative experiences with accommodations at work), and personal factors (e.g., stress).

In terms of method, Silveira et al.~\cite{silveira19} analyzed how gender relates to confidence in programming in the 2018 SO survey.
They find that women and non-binary respondents believe they are not as good as their peers.

\section{Methodology}


Stack Overflow is an online Question and Answer (Q\&A) platform that provides individuals with an opportunity to discuss a wide range of topics \cite{barua2014developers}. 
Starting 2011, Stack Overflow has been conducting an annual survey\footnote{\url{https://survey.stackoverflow.co/}} to understand the user demographics and their practices. The results and the raw data of the annual developer surveys are available publicly and can be analyzed to gain deeper insights. 
We took this survey as a starting point for our study, as we describe in the following.

\subsection{Data Collection: The 2022 Stack Overflow Developer Survey}
The 2022 annual Stack Overflow Developer Survey\footnote{\url{https://survey.stackoverflow.co/2022}} was conducted from May 11, 2022 to June 1, 2022. A total of 73,268 respondents participated in this developer survey, of whom 70.03\% were developers by profession. According to the information available on the Stack Overflow website, the survey respondents were recruited with the help of channels owned by Stack Overflow such as blog posts, onsite messaging, or email lists.
While Stack Overflow does not define what their target population is, they name the survey the ``Developer'' survey, and talk about ``software developers'' in their methodology. 
%

The survey questions can be categorized into different blocks, e.g., asking about demographics (e.g., age, company size, country), used tools and technologies, use of Stack Overflow, and productivity at work. 
In addition, for the first and last time since 2011, the 2022 survey collected information about mental health, including a question about neurodiversity and/or any emotional or anxiety disorder.

The 2022 survey also included seven knowledge questions (K1-K7) and three frequency questions (F1-F3). The knowledge questions focused on the respondents' ability to find knowledge and information within their organization. The respondents had to answer these knowledge questions on a five-point Likert scale ranging from `strongly disagree' to `strongly agree'. The frequency questions focused on how frequently the respondents experienced situations in which they had to interact with others in their organization and encountered productivity difficulties. The responses to these three frequency questions were also obtained on a five-point Likert scale, ranging from `never' to `10+ times a week'. The questions are as follows.
\begin{itemize}
    \item K1: \lq I have interactions with people outside of my immediate team \rq
    \item K2: \lq Knowledge silos prevent me from getting ideas across the organization (i.e., one individual or team has information that isn't shared with others)\rq 
    \item K3: \lq I can find up-to-date information within my organization to help me do my job. \rq
    \item K4: \lq I am able to quickly find answers to my questions with existing tools and resources.\rq 
    \item K5: \lq I know which system or resource to use to find information and answers to questions I have.\rq 
    \item K6: \lq I often find myself answering questions that I've already answered before \rq 
    \item K7: \lq Waiting on answers to questions often causes interruptions and disrupts my workflow \rq
    \item F1: \lq Needing help from people outside of your immediate team? \rq
    \item F2: \lq Interacting with people outside of your immediate team? \rq 
    \item F3: \lq Encountering knowledge silos (where one individual or team has information that's not shared or distributed with other individuals or teams) at work\rq 
\end{itemize}

Since existing work in SE and in general workplace studies indicates that ND and NT individuals might differ in their ability to find information at work and their interaction with others, we chose to analyze the differences between the responses of ND and NT respondents to these ten questions.
Thus, in this work, we (i) descriptively analyze the respondent demographics and the time spent answering / searching for questions and (ii) investigate whether there exist statistical differences between the responses to the seven knowledge questions and the three frequency questions between ND and NT individuals. 


\subsubsection{Filtering the 2022 Stack Overflow Developer Survey} 
We only considered respondents who were full-time employed and over 18 years of age. 
Since we were interested in investigating the differences (if any) between the group of ND individuals and the group of NT individuals, we further filtered the data into different groups, i.e., those having (i) dyslexia without any co-morbidity (i.e., having learning differences), (ii) ADHD without any co-morbidity, (iii) ASD without any co-morbidity and (iv) no neurodiversity (none of the specified conditions). 
We removed responses where the first knowledge question (K1) and the first frequency question (F1) were not answered, as most of these answers had skipped this part of the survey altogether.
This resulted in 363 individuals in the dyslexic group, 1305 individuals in the ADHD group and 374 individuals in the ASD group. For each of these three groups, we randomly selected an equal number of NT individuals.
We then removed these individuals from the pool of NT individuals for the remaining sampling steps, ensuring that no NT individual is present in more than one sample.

Since the responses to the 7 knowledge questions and 3 frequency questions can be impacted by the work mode as well as the number of years of work experience one has, we additionally adopted a paired sampling approach based on (a) work mode and (b) number of years of work experience. We filtered the data to include datasets of dyslexic, ADHD and ASD individuals working (i) fully remote, (ii) full in-person and (iii) hybrid. This resulted in 148 dyslexic, 664 ADHD and 176 ASD individuals working fully remote; 34 dyslexic, 133 ADHD and 41 ASD individuals working full in-person; and 181 dyslexic, 507 ADHD and 157 ASD individuals working hybrid. For each of the ND groups belonging to the three different work modes, we randomly selected an equal number of NT individuals, still ensuring that no NT individual is present in more than one sample. 
We performed the same kind of sampling for work experience, by grouping participants into 0-5 years of work experience, 6-10 years of work experience, and 10+ years of work experience.

\subsection{Data Analysis}
We started our analysis by calculating descriptive statistics for questions addressing time spent for answering questions and for searching for solutions to problems. 
To analyze the various countries reported by survey participants, we grouped them by continent.
To do so, we use the division of National Geographic\footnote{\url{https://education.nationalgeographic.org/resource/Continent/}} into seven continents, i.e., Asia, Africa, North America, South America, Antarctica, Europe, and Australia, as well as ``Other'' and ``Nomadic''.
%
%

To answer our RQ (``What differences exist between professional developers with dyslexia, ADHD, and ASD and their neurotypical peers?''), we test whether there exist statistically significant differences in the responses to K1-K7 and F1-F3 between the ND and NT groups.
To do so, we carried out non-parametric Mann-Whitney U tests. 
Since the individual participants in the Stack Overflow survey are independent, the key assumption of the Mann-Whitney U test regarding independence is given.
%
We used an alpha level of 5\% for all tests.
While we ensured unique NT groups for each comparison, there is an overlap within the groups each condition (i.e., Dyslexic, ADHD and ASD) for different conditions. For instance, a dyslexic individual working fully remote is also present in the overall sample of dyslexic individuals and in one of the work experience samples.
Thus, due to multiple Mann-Whitney U tests being conducted on data groups that might have common instances, the probability of making Type I errors (family-wise errors) increases \cite{field2012discoveringstats}. To control for this probability, we applied the Bonferroni-Holm correction~\cite{holm1979simple} on all tests within one condition (i.e., dyslexia, ADHD, and ASD). 
%

\subsection{Validity Threats}
In the following, we will discuss potential threats to the validity of our study in terms of internal, external, and construct validity, as well as reliability.

\subsubsection{Internal Validity}
While pairing the ND and NT groups based on number of years of work experience, we consider range in the years of work experience and do not match based on same mean years of experience. 
To adjust for inflated Type I errors, we used the Bonferroni-Holm correction for each of the ND groups.
This might lead to overly conservative results, i.e., there might be more differences between the ND and NT groups.
 
The Stack Overflow developer survey was conducted during COVID-19, which could affect the effect of remote work.
However, we argue that it might indeed be the most suitable timing for this survey, as software developers might have been working according to their preferred work mode, not restricted by either remote work or return-to-office mandates.

We selected ND individuals who had no other ND or mental health condition.
Apart from filtering out individuals with multiple ND conditions, such as a combination of ASD and ADHD, this also included filtering out people with mental health issues such as anxiety or depression.
While this strengthens the chance that observed differences are in fact due to the ND condition, it might also exclude more severe cases, e.g., where a ND condition has already led to developing anxiety or depression, or where the symptoms of multiple ND conditions cause more challenges.
%
%

%
%

 
\subsubsection{External Validity}
It is likely that respondents who took part in this survey are not representative of the ND and NT developers worldwide. While this poses a threat to external validity, we also observe that there is a relatively broad representation across different continents in comparison to many surveys in SE. This is further discussed in Section~\ref{sec:descriptive}.
Additionally, the focus of the Stack Overflow survey is on developers and might limit the generalizability to other SE-related roles.

\subsubsection{Construct Validity}
Our aim is to understand the differences and/or similarity in the experiences of ND developers from that of NT developers in the SE domain. Though knowledge sharing, interaction, and finding information play a key role in determining one’s experience at the workplace, there can be other factors that can contribute to the overall work experience and satisfaction. 
%
%
%

It is possible that the questions used in the survey are worded in an ambiguous way. Additionally, they might not measure valid constructs, thereby influencing the responses obtained. 
Since we did not design the survey, we do not have much information about its theoretical framework and had no influence on the design and wording of the questions. 

\subsubsection{Reliability}
The dataset of the 2022 Stack Overflow developer survey as well as the questionnaire are publicly available under \url{https://survey.stackoverflow.co/2022}. In addition, we published the filtered data we used, our final samples, as well as the scripts for sampling and analysis \cite{dataset_SO}, so that others can replicate our work or use it as an inspiration for future work.

\section{Results}
In the following, we describe our results.
We start with an overview of descriptive statistics of the demographic data of our unpaired samples (i.e., all ND individuals and the random corresponding samples of NT individuals).
Similar descriptive statistics can be obtained for the paired samples using the scripts published as a part of our dataset.
Then, we present the results of the statistical tests, answering \textbf{RQ}.

\subsection{Demographics and Descriptive Statistics}
\label{sec:descriptive}
%
We observed that, for all three conditions, there is a clear difference between the origins of the survey respondents. For individuals with Dyslexia and the corresponding NT group, the majority of the respondents were from Europe (63.36\% and 37.74\% respectively). A similar trend was observed for individuals with ASD and the corresponding NT group, with the majority of the respondents being from Europe (52.94\% and 42.51\% respectively). For individuals with ADHD, the majority of the respondents were from North America (45.67\%) with majority of the respondents in the corresponding NT group being from Europe (40.08\%).

In terms of education, all three samples are relatively similar to their NT counterparts, with the majority in all samples having Bachelor degrees (51.49\% of individuals with ADHD, compared to 51.84\% in the NT sample; 44.65\% of individuals with ASD, compared to 47.33\% in the NT sample; 44.35\% of individuals with dyslexia, compared to 48.21\% in the NT sample). 
It is noteworthy that fewer ND individuals have a Master degree (15.71\% of individuals with ADHD, compared to 25.46\% in the NT sample; 25.40\% of individuals with ASD, compared to 28.88\% in the NT sample; 25.34\% of individuals with dyslexia, compared to 26.72\% in the NT sample) and more have attended college/university without earning a degree (17.24\% of individuals with ADHD, compared to 11.04\% in the NT sample; 15.24\% of individuals with ASD, compared to 10.43\% in the NT sample; 13.77\% of individuals with dyslexia, compared to 12.12\% in the NT sample).

In terms of coding experience, individuals with ADHD indicated an average of 12 years of coding experience, with their NT peers reporting an average of 13 years. 

The majority of respondents were within the 25-34 age range. Among individuals with dyslexia, 51.24\% (N = 186) were in this category, compared to 51.52\% (N = 187) of NT. For individuals with ASD, 45.72\% (N = 171) were within this age range, lower than the 52.94\% (N = 198) of their NT counterparts. In the ADHD group, 52.57\% (N = 686) were aged 25-34, compared to 49.66\% (N = 648) among NT.

\begin{figure}[ht]
\includegraphics[width=8cm]{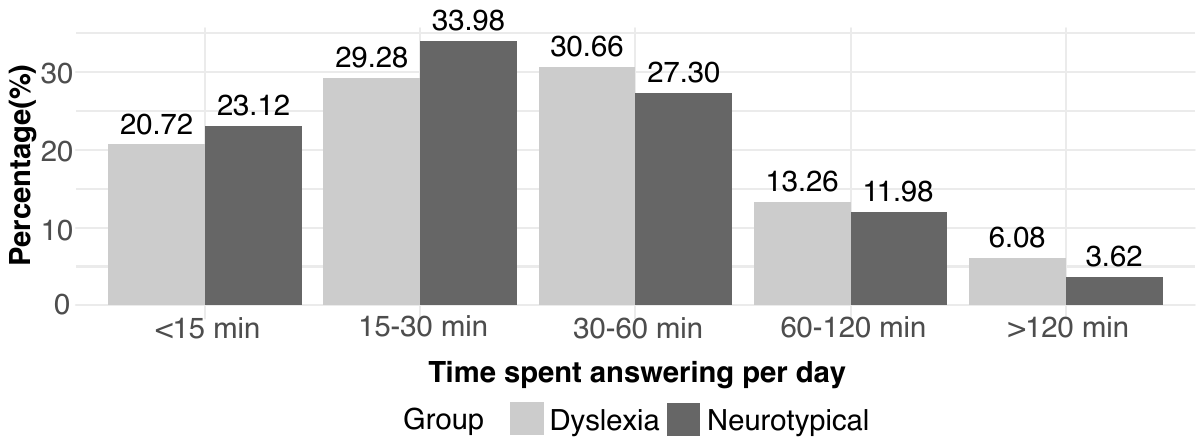}
\caption{Daily time answering questions in the Dyslexia sample and the corresponding neurotypical sample.}
\label{fig:timeAns_dys}
\end{figure}

\begin{figure}[ht]
\includegraphics[width=8cm]{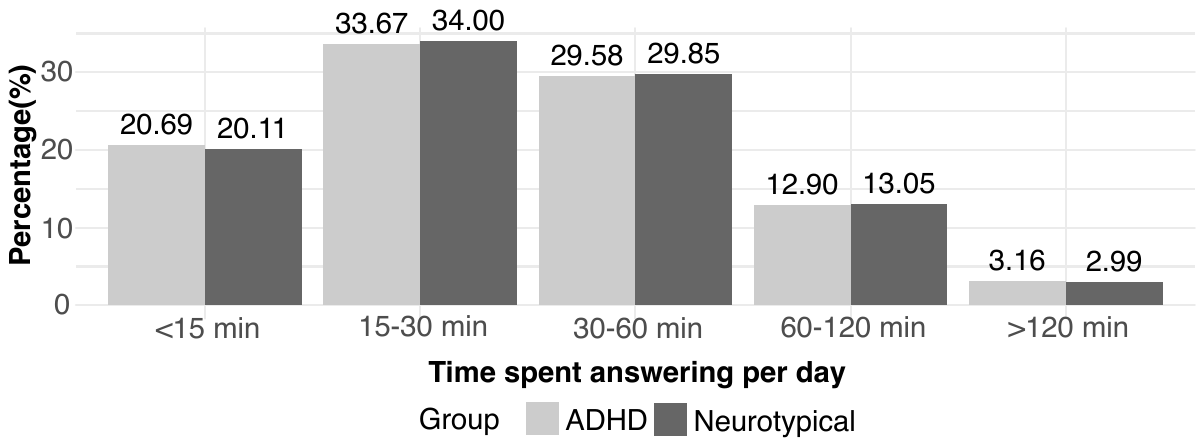}
\caption{Daily time answering questions in the ADHD sample and the corresponding neurotypical sample.}
\label{fig:timeAns_adhd}
\end{figure}

\begin{figure}[ht]
\includegraphics[width=8cm]{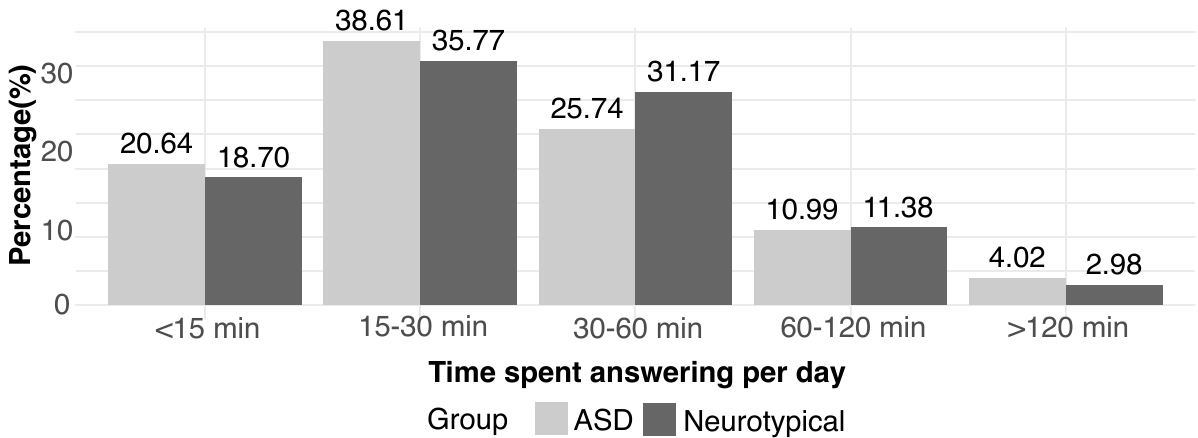}
\caption{Daily time answering questions in the ASD sample and the corresponding neurotypical sample.}
\label{fig:timeAns_asd}
\end{figure}

Figures~\ref{fig:timeAns_dys}, \ref{fig:timeAns_adhd}, and \ref{fig:timeAns_asd} show the comparisons of time spent in answering work-related questions by individuals with dyslexia, ADHD, and ASD. 
Most individuals across all groups spent between 15-60 minutes per day.
Interestingly, the time individuals with ADHD spent on answering questions is almost identical to their NT counterparts, while there are more differences observable in the other two groups.

\begin{figure}[ht]
\includegraphics[width=8cm]{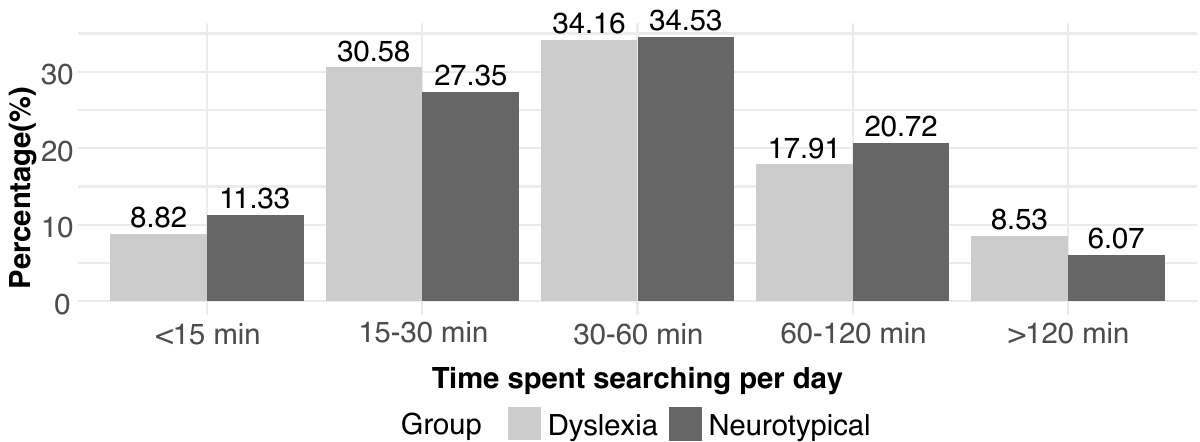}
\caption{Daily time searching in the Dyslexia sample and the corresponding neurotypical sample.}
\label{fig:timeSearch_dys}
\end{figure}

\begin{figure}[ht]
\includegraphics[width=8cm]{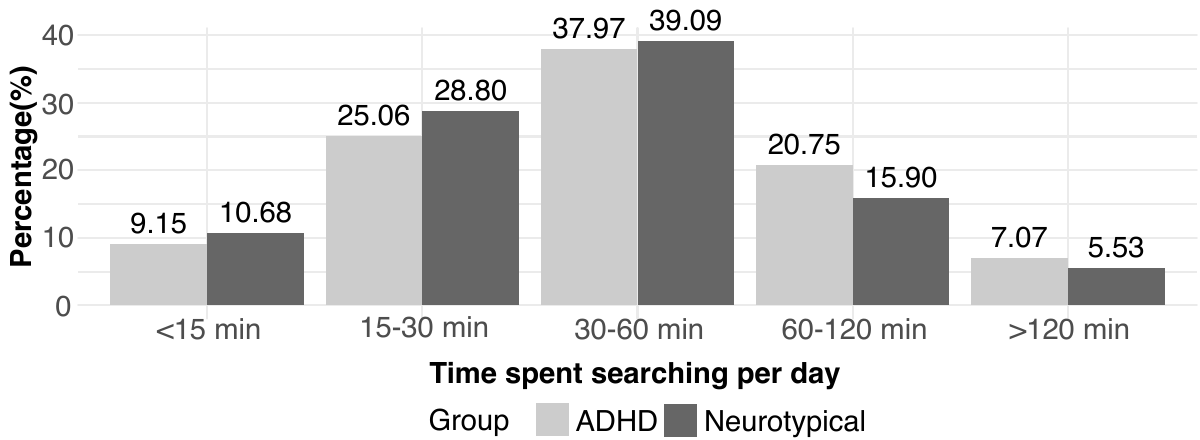}
\caption{Daily time searching in the ADHD sample and the corresponding neurotypical sample.}
\label{fig:timeSearch_adhd}
\end{figure}

\begin{figure}[ht]
\includegraphics[width=8cm]{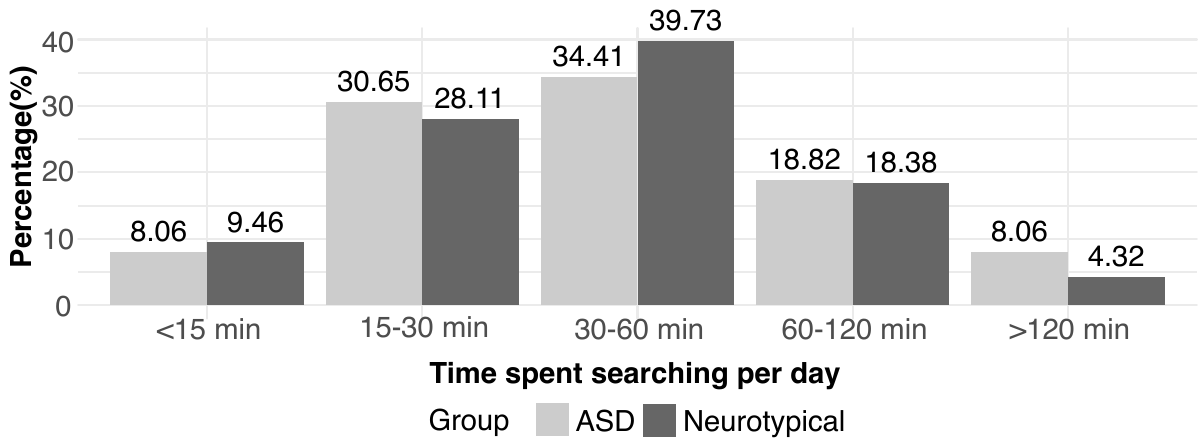}
\caption{Daily time searching in the ASD sample and the corresponding neurotypical sample.}
\label{fig:timeSearch_asd}
\end{figure}

Figures~\ref{fig:timeSearch_dys}, \ref{fig:timeSearch_adhd}, and \ref{fig:timeSearch_asd} show the comparisons of time spent searching for answers or solutions to work-related problems by individuals with dyslexia, ADHD, and ASD. 
For all three conditions, there seems to be a difference to NT peers in terms of individuals spending more than 120 minutes per day.

\subsection{Inter-group Comparison for Unpaired Sampling}
We first compare the responses to K1-K7 and F1-F3 using an unpaired sampling approach. We then check for differences in these 10 questions using a paired sampling approach where we pair ND and NT individuals based on (i) work mode and (ii) number of years of experience. 
Table~\ref{tab:my_label} summarizes the results of Mann-Whitney U Test with Bonferroni-Holm correction for the 10 questions (K1-K7 and F1-F3) for the ND groups and the corresponding randomly selected NT group. \\
\begin{table}[ht!]
    \caption{Corrected p-value for neurodivergent vs. randomly sampled neurotypical groups}
    \label{tab:my_label}
    \centering
    \begin{threeparttable}
    \begin{tabular}{| l | l | l | l |}
    \cline{2-4}
    \multicolumn{1}{c|}{}&  Dyslexia vs. NT & ADHD vs. NT & Autism vs. NT  \\ \hline
    K1  & {1.00} & {1.00}& {1.00} \\ \hline
    K2  & {1.00} & {1.00}& {1.00} \\ \hline
    K3  & {1.00} & {1.00} & {1.00} \\ \hline
    K4  & {1.00} & {0.989} & \textcolor{red}{\textbf{0.003**}} (-0.148) \\ \hline
    K5  & {1.00} & {0.076} & {1.00} \\ \hline
    K6 & {1.00} & {1.00}& {0.285} \\ \hline
    K7  & {1.00} & \textcolor{red}{\textbf{0.0006**}} (-0.086) &  {1.00} \\ \hline
    F1  & {1.00} & {1.00} & {1.00} \\ \hline
    F2  & {1.00} & \textcolor{red}{\textbf{0.020*}} (-0.070) & {1.00} \\ \hline
    F3  & {1.00} & {0.254} &  {1.00} \\ \hline
    \end{tabular}
\begin{tablenotes}
   \item Note: * indicates significant difference at p-value $< 0.05$ \\
   ** indicates significant difference at p-value $< 0.01$ \\
   value in () indicates the effect size \\
   NT represents randomly selected Neurotypical group
\end{tablenotes}
    \end{threeparttable}
\end{table}
For individuals with ADHD and the randomly selected NT group, we observe statistically significant differences (with a small effect size) for K7 and F2 respectively, i.e., the ADHD group faces interruptions in their workflow while waiting for answers more often, and they have fewer interactions outside their team.
For individuals with ASD, we observe a statistically significant difference in K4 (with a small effect size).
That is, individuals with ASD have a harder time finding answers to their questions with existing tools and resources. 

\subsection{Inter-group Comparison for Paired Sampling Based on Work Mode}
Since it is possible that experience can vary based on whether they work fully remote, in-person or hybrid, it is important to understand if differences in the responses to the knowledge and frequency questions exist when the groups are matched on the basis of work mode. 

The results of Mann-Whitney U Test with Bonferroni-Holm correction for the ND versus NT groups working fully remote and fully in person resulted in no statistically significant differences for any of the ten questions (K1-K7 and F1-F3). 

Table~\ref{tab:my_label3} summarizes the results of Mann-Whitney U Test with Bonferroni-Holm correction for the 10 questions (K1-K7 and F1-F3) for the ND versus NT groups working in hybrid mode.

\begin{table}[ht!]
    \caption{Corrected p-value for Neurodivergent vs. Neurotypical groups working in hybrid mode}
    \label{tab:my_label3}
    \centering
    \begin{threeparttable}
    \begin {tabular}{| l | l | l | l |}
    \cline{2-4}
    \multicolumn{1}{c|}{}&  Dyslexia vs. NT & ADHD vs. NT & Autism vs. NT  \\ \hline
    K1  & {1.00} & {1.00}& {0.417} \\ \hline
    K2  & {1.00} & {1.00}& {1.00} \\ \hline
    K3  & {1.00} & {1.00}& {0.051} \\ \hline
    K4  & {0.565} & {0.474} & {0.055}  \\ \hline
    K5  & {0.569} & {1.00}& {0.979} \\ \hline
    K6 & {1.00} & {1.00} & {1.00} \\ \hline
    K7  & {1.00} & \textcolor{red}{\textbf{0.0001**}} (-0.147) & {1.00} \\ \hline
    F1  & {1.00} & {1.00}& {1.00} \\ \hline
    F2  & {1.00} & \textcolor{red}{\textbf{0.0003**}} (-0.142) & {1.00} \\ \hline
    F3  & {1.00} & {1.00}& {1.00} \\ \hline
    \end {tabular}
\begin{tablenotes}
    \item Note: ** indicates significant difference at p-value $< 0.01$ \\
    value in () indicates the effect size \\
    NT represents randomly selected Neurotypical group
\end{tablenotes}
    \end{threeparttable}
\end{table}
Our results indicate that there are no significant differences between the responses for individuals with dyslexia and their NT counterparts and between individuals with ASD and the corresponding NT individuals working in hybrid mode. However, similar to our finding for the unpaired samples, we found statistically significant differences (with small effect size) for individuals with ADHD in questions K7 and F2.
Hence, individuals with ADHD working in hybrid mode face interruptions in their workflow while waiting for answers more commonly than their NT peers. Similarly, they have fewer interactions outside their team.

\subsection{Inter-group Comparison for Paired Sampling Based on Number of Years of Work Experience}
In addition to work mode, work experience could have an effect on the answers. Thus, we tested differences in the responses to the knowledge and frequency questions (K1-K7 and F1-F3) when pairing groups on the basis of number of years of work experience. For this, we conducted inter-group comparisons using Mann-Whitney U tests along with the Bonferroni-Holm correction while matching the ND and NT groups based on their number of years of work experience.

\subsubsection{0-5 Years} 
Table~\ref{tab:my_label4} summarizes the results of the Mann-Whitney U tests with Bonferroni-Holm correction for the 10 questions (K1-K7 and F1-F3) for the ND versus NT groups with 0 to 5 years of work experience.

\begin{table}[ht!]
    \caption{Corrected p-value for Neurodivergent vs. Neurotypical groups having 0-5 years of work experience}
    \label{tab:my_label4}
    \centering 
    \begin{threeparttable}
    \begin {tabular}{| l | l | l | l |}
    \cline{2-4}
    \multicolumn{1}{c|}{}&  Dyslexia vs. NT & ADHD vs. NT & Autism vs. NT  \\ \hline
    K1  & {1.00} & {1.00}& {1.00} \\ \hline
    K2  & {1.00} & {1.00}& {1.00} \\ \hline
    K3  & {1.00} & {1.00}& {1.00} \\ \hline
    K4  & {1.00} & {1.00}& {1.00} \\ \hline
    K5  & \textcolor{red}{\textbf{0.008**}} (-0.219) & {1.00}& {1.00} \\ \hline
    K6 & {1.00} & {1.00}& {1.00}\\ \hline
    K7  & {1.00} & {1.00} & {1.00} \\ \hline
    F1  & {1.00} & {1.00}& {1.00} \\ \hline
    F2  & {1.00} & {1.00}& {1.00} \\ \hline
    F3  & {1.00} & {1.00} & {1.00} \\ \hline
    \end {tabular}
\begin{tablenotes}
    \item Note: ** indicates significant difference at p-value $< 0.01$ \\
    value in () indicates the effect size \\
    NT represents randomly selected Neurotypical group
\end{tablenotes}
    \end{threeparttable}
\end{table}
We found a statistically significant difference for K5 (with medium effect size) between responses from individuals with dyslexia and the randomly selected NT individuals. None of the other comparisons between the three ND groups and their corresponding NT groups was significant.

\subsubsection{6-10 Years} 
The results of the Mann-Whitney U tests with Bonferroni-Holm correction for the ND versus NT groups with 6 to 10 years of work experience indicated no statistical significance for the 10 questions (K1-K7 and F1-F3).


\subsubsection{10+ Years} 
Similar to our comparison for 6-10 years of work experience, we observed no statistical significance for the 10 questions (K1-K7 and F1-F3) for the ND versus NT groups with 10+ years of work experience.

\section{Discussion}
Our results show few significant differences between ND and NT developers.
This could indicate that other factors, e.g., the type of work they do or their seniority, have a larger influence on knowledge sharing, interaction, and finding information.
Similarly, the severity of ND conditions, as well as potentially undisclosed or undiagnosed conditions in our NT sample could affect the results.
This is particularly noteworthy given the differences in the ND paradigm across cultures. Hirota et al.~\cite{hirota2024neurodiversityculture} report that, in many cultures, uniqueness due to ND could be interpreted negatively thereby affecting interpersonal relations. This could lead to individuals in such cultures not seeking diagnosis due to fear of discrimination or stigmatization, as also highlighted in the SE context in \cite{liebel24_adhd}. 
To study this in more depth, and to allow for generalizable results, replication studies are needed with more systematic sampling.

Our results show differences in responses to two questions (i.e., K7 and F2) between developers with ADHD and NT developers working in hybrid mode. However, we did not observe any difference in any of the questions between the three groups of ND developers and the corresponding NT groups working fully remote or fully in-person. As mentioned previously, the survey was conducted towards the end of COVID-19, at a time when developers could likely work in their preferred work mode and not as restricted by their organizations. 
Ralph et al.~\cite{ralph2020pandemic} find that the COVID-19 pandemic had negative effects on most developers, especially on well-being and productivity.
While the authors note that people with disabilities are disproportionally affected, they do not specifically mention ND developers.
Furthermore, this disproportionate effect could manifest in unexpected ways.
For instance, while ND developers might have been affected stronger with respect to well-being, working from home could indeed have helped them in other ways, e.g., in reducing distractions or allowing for more suitable interaction and communication styles with their teams.
Indeed, existing qualitative study highlight that ND software engineers often like to work from home, in an environment tailored to their specific needs \cite{morris2015understanding,liebel24_adhd}.
Similarly, ND developers choosing to work fully in person may be those with less severe conditions, or NT developers working from home may be those with challenges similar to ND peers, including with undiagnosed ND conditions.
Further investigations that take into account such information might reveal meaningful insights about the differences in the workplace experiences of ND and NT engineers working in different work modes and can be helpful in creating more inclusive workplace environments. 

Since we did not conduct the survey ourselves, we had no influence on the questions asked.
As a result, there are several questions that might suffer from low validity, and aspects of ND challenges that remain unexplored in this study.
Regarding the first point, it is for example unclear whether the concept of ``interaction'' used in two questions (i.e., K1 and F2) relates only to work-related interaction or also describes social interaction. Further, cultural differences can affect workplace interaction. For instance, Sanchez-Burks et al. \cite{sanchez2009culturaliteraction} report that individuals from different cultural backgrounds often use ``different relational schemas to navigate their workplace interactions". Thus, differences in the cultural background of the respondents might have had an impact on their interpretation of questions K1, F1 and F2.

A third limitation of this study is that we included only employed and retired developers.
However, studies on neurodiversity at the workplace show large percentages of unemployment among ND individuals, e.g., up to 50-75\% of individuals with ASD \cite{hendricks2010employment}.
As these individuals are likely to face more challenges than employed individuals, considering the perspective of unemployed ND software engineers could show more pronounced patterns.

Despite the limitations of the Stack Overflow survey and our analysis, this study provides a good baseline for future work.
For instance, as hybrid work is common in SE, the findings that developers with ADHD working in hybrid mode experience more interruptions when waiting on answers compared to their NT peers warrants further investigation of this setting.
Similarly, while it is expected that the demographics of our comparison groups differ, we observe similar patterns in many of the demographic questions, as reported in Section~\ref{sec:descriptive}.
For instance, organization sizes and the time spent answering questions and searching for information are similar in the ND groups and the unpaired random NT samples.
One exception is the global distribution, where the continents differ clearly.

Our findings support that ND conditions affect the individuals to some extent, even in environments that they might tailor to their needs (e.g., remote or hybrid).
This connects well to existing work, e.g., by Gama et al.~\cite{liebel24_adhd,gama25_gt_asd_adhd} and Morris et al.~\cite{morris2015understanding}.
However, the survey lacks details on important aspects that could affect their challenges.
Specifically, both Gama et al.~\cite{gama25_gt_asd_adhd} and Morris et al.~\cite{morris2015understanding} highlight the important effect of accommodations and awareness, and the role that the environment plays for ND individuals - both aspects that are not visible in the Stack Overflow survey.


While differences are small, there is a larger percentage of ND developers spending more than 120 minutes per day answering questions and searching for information than NT individuals. This could relate to common symptoms in all conditions related to attention, memory, and reading and writing skills.


\section{Conclusion}
In this paper, we present an analysis of the 2022 Stack Overflow developer survey with respect to neurodiversity, answering the research question ``What differences exist between professional developers with dyslexia, ADHD, and ASD and their neurotypical peers?''.
We quantitatively compare the answers of professional developers with dyslexia (n=363), ADHD (n=1305), and ASD (n=374) with randomly sampled neurotypical professionals in terms of knowledge sharing, interaction, and finding information.
We also compare the responses of the neurodivergent developers with neurotypical developers paired on the basis of work mode and number of years of work experience. 

Our findings show few significant differences between the groups. 
Among the observed differences, we find that developers working with ADHD working in a hybrid work mode face more difficulties in obtaining and sharing knowledge within their organization.
Despite these few differences, our current work lays a strong foundation for future research in this domain to understand the needs and experiences of neurodivergent software engineers with an overall aim of creating more inclusive workplace environment.

Various aspects are not considered or potentially invalid in the Stack Overflow survey, and might influence the results, e.g., the impact of accommodations at the workplace and tailoring of remote or hybrid work environments.
On the one hand, this means there exist various threats to validity in our findings.
On the other hand, this also means there is ample potential for follow-up work in the future.
For instance, taking a closer look at cultural differences is an important next step, as our samples differ in the reported continents.
Similar, many of the strengths and difficulties of neurodivergent software engineers reported in existing SE work, such as creativity, systems thinking, or difficulties in keeping up attention, are not covered by the Stack Overflow Developer Surveys.
These points should therefore be investigated in the future.

As such, our study likely provides a conservative baseline that can be used in future studies to further explore the challenges of ND professionals in SE, as well as how to address them.



\end{document}